\documentclass[twocolumn,superscriptaddress,nofootinbib,notitlepage,longbibliography,aps]{revtex4-1}
\pdfoutput=1

\usepackage{graphicx}
\usepackage{amsmath}
\usepackage{amssymb}
\usepackage{amsfonts}
\usepackage[dvipsnames]{xcolor}
\usepackage[linktoc=none]{hyperref}
\usepackage[utf8]{inputenc}
\usepackage{comment}

\hypersetup{colorlinks=true,linkcolor=blue,citecolor=teal,filecolor=magenta, urlcolor=cyan}
\renewcommand{\arraystretch}{1.5}

\newcommand{\UNINA}{Dipartimento di Fisica ``Ettore Pancini'', Università degli studi di Napoli ``Federico II'', Complesso Universitario Monte S. Angelo, I-80126 Napoli, Italy}
\newcommand{\INFN}{INFN - Sezione di Napoli, Complesso Universitario Monte S. Angelo, I-80126 Napoli, Italy}
\newcommand{\SSM}{Scuola Superiore Meridionale, Università degli studi di Napoli ``Federico II'', Largo San Marcellino 10, 80138 Napoli, Italy}

\begin{document}

\title{A strike of luck: could the KM3--230213A event \\be caused by an evaporating primordial black hole?}

\author{Andrea Boccia}
\affiliation{\SSM}
\affiliation{\INFN}
\author{Fabio Iocco}
\affiliation{\UNINA}
\affiliation{\INFN}

\begin{abstract}
We investigate whether the ultra high energy neutrino inferred by the recent KM3NeT observation could have originated from an evaporating black hole. Given the characteristics of black hole (BH) evaporation mechanism, any object capable of producing particles in the energy range of the detected event ($\sim$100-800 PeV) must have a mass $M_{\rm BH} \lesssim 10^7 \rm g$. No known astrophysical mechanism can generate black holes of such low mass, leaving primordial black holes (PBHs) --potentially formed at the end of cosmic inflation-- as the only viable candidates. Black holes with masses $M_{\rm BH} \lesssim 10^7 \rm g$ have lifetimes shorter than $10^{-5}$ seconds, meaning PBHs in this mass range should have fully evaporated by now. However, recent studies suggest that quantum effects, collectively referred to as the “memory burden”, may slow down black hole evaporation, potentially extending the lifetimes of low--mass PBHs to timescales comparable to or exceeding the Hubble time.
We systematically explore the parameter space of memory-burdened PBHs, assuming that they constitute a fraction of the dark matter ($f_{\rm PBH}$) within current constraints, and identify viable regions that could explain the KM3–230213A event. We further predict the occurrence rate of similar events, and find that KM3NeT and future neutrino experiments could test this scenario over the next years.
\end{abstract}

\maketitle

\par{\bf Introduction} The KM3NeT collaboration has recently claimed the observation of one muon event with an energy of $120^{+110}_{-60}$ PeV (event KM3--230213A) probably originated from a neutrino of energy $E_{\nu}\sim 110-790 \rm \, PeV$ (90 \% C.L.)  \cite{KM3NeT:2025npi}.

The neutrino with the highest energy ever recorded, its origin is yet unknown and highly debated. Possible explanations range from Pevatrons of extragalactic origins \cite{KM3NeT:2025bxl,Dzhatdoev:2025sdi,Neronov:2025jfj} to cosmogenic neutrinos \cite{KM3NeT:2025vut}.

\par It is reasonable to ask whether a neutrino of such energy may have been emitted by an evaporating black hole.
Owing to the Hawking evaporation mechanism, black holes are thought to emit a spectrum of all particles, with characteristic energy increasing for decreasing mass.
In order to produce particles of the energy observed by the KM3NeT collaboration, the mass of the emitting black hole should be $M_{\rm BH} \lesssim 10^7 \rm g$.

If one wants to invoke the evaporation of a black hole (BH) of such mass to explain the KM3--230213A event, one must reckon with two critical issues: 
{\it a)} no mechanism involving known astrophysics is able to generate a black hole of such low mass;
{\it b)} due to the evaporation mechanism characteristics, black holes of masses $M_{\rm BH} \lesssim 10^7 \rm g$ have a lifetime shorter than $10^{-5}$ seconds. A viable opposition to point {\it (a)} is that a class of primordial black holes (PBH) may have been created by the direct collapse of density anisotropies in the early Universe ~\cite{Zeldovich:1967lct, Hawking:1971ei, Carr:1974nx}, and could constitute all or part of the Dark Matter ~\cite{Carr:2016drx, Green:2020jor, Carr:2020gox, Carr:2021bzv}.
Due to their short lifetimes ({\it b}), PBHs of masses capable of emitting at the PeV should, however, have evaporated by now.\\

The standard semi-classical picture of black hole evaporation assumes that the object remains classical throughout its lifetime~\cite{Hawking:1975vcx}. However, growing evidence suggests that this framework may be incomplete, particularly in addressing the information loss paradox~\cite{Preskill:1992tc}. A key limitation of Hawking’s derivation is the neglect of back-reaction effects, how the emission of radiation influences the quantum state of the black hole itself. This omission becomes critical when the energy of emitted particles approaches the total energy of the black hole.

Recent research~\cite{Dvali:2018xpy, Dvali:2020wft, Dvali:2024hsb,Davies:2024ysj} suggests that such back-reaction effects give rise to a universal “memory burden”. This phenomenon emerges because the information encoded in a system resists its decay, driven by the response of quantum modes tied to its entropic degrees of freedom. As a result, once a black hole’s mass drops below a certain threshold, back-reaction significantly alters its evolution, slowing the evaporation process and extending its lifetime. This opens the possibility that black holes with masses $M_{\rm PBH} \lesssim 10^{15}$ g may still be evaporating today, with potential observational consequences~\cite{Balaji:2024hpu, Barman:2024iht, Bhaumik:2024qzd, Barman:2024ufm, Kohri:2024qpd, Jiang:2024aju}.

Furthermore, the memory burden effect challenges conventional constraints on the existence of light PBHs, suggesting that even those with $M_{\rm PBH} \lesssim 10^9$ grams could persist as viable dark matter candidates~\cite{Dvali:2020wft, Dvali:2021byy, Alexandre:2024nuo, Thoss:2024hsr, Haque:2024eyh}.

In the following we first recall the fundamental equations for BH evaporation and memory burden mechanism. We then describe the method used to determine the number of events expected inside the KM3NeT instrument, and to select the PBH parameters combinations capable of producing such an event today. For the latter, we finally show forecasts of the expected future event counts within the KM3NeT detector.

\par{\bf Black Hole evaporation formalism} Quantum effects predict that primordial black holes (PBHs) should emit radiation, a phenomenon originally proposed by Hawking \cite{Hawking:1975vcx}. This emission follows an approximately thermal spectrum, peaking at the characteristic Hawking temperature
\begin{equation}
\label{eq:hawktemp}
T_{\rm H} = \frac{1}{8 \pi G M_{\rm PBH}} \simeq 10^{4}\left(\frac{10^9~{\rm g}}{M_{\rm PBH}}\right){\rm GeV},
\end{equation}
where  $G$  is the gravitational constant, and $M_{\rm PBH}$ is the black hole mass expressed in grams. Since this radiation originates from the PBH’s gravitational energy, its mass decreases over time, with the loss rate given by
\begin{equation}
\label{eq:mdot}
\frac{{\rm d}M_{\rm PBH}}{{\rm d}t} = - \frac{\mathcal{G} \, g_{\rm H} }{30720 \pi \, G^2 M_{\rm PBH}^2}.
\end{equation}
Here,  $\mathcal{G} \sim 3.8 $ \cite{Page:1976wx} accounts for back-scattering effects, while  $g_{\rm H} \sim 102.6 $ \cite{Mazde:2022sdx} corresponds to the spin-weighted number of degrees of freedom at temperature $T_{\rm H}$ . According to the standard evaporation scenario, this process continues until the PBH completely vanishes. The time until full evaporation is given by
\begin{equation}
\label{eq:taubh}
\tau_{\rm PBH} = \frac{10240 \pi G^2 M_{\rm PBH}^3}{\mathcal{G} g_H} \simeq 4.4\times 10^{17}\left(\frac{M_{\rm PBH}}{10^{15}{\rm g}}\right)^3~{\rm s}.
\end{equation}
As a result, PBHs with \( M_{\rm PBH} \lesssim 10^{15} \) g would have already evaporated, given the current age of the Universe.\\ The spectra of the emitted quanta will be thermal-like 
\begin{equation}
    \label{eq:hawkspec}
    \frac{{\rm d}^2N_i}{{\rm d}E{\rm d}t} = \frac{g_i}{2 \pi} \frac{\mathcal{F}(E,M_{\rm PBH})}{e^{E/T_{\rm H}} + 1} \, ,
\end{equation}
with $g_i $ being the number of internal degrees of freedom of the emitted particle, and $\mathcal{F}(E,M_{\rm PBH})$ the gray-body factor.\\
If  memory burden effect is taken into account, the evaporation process deviates from this classical picture. The PBH evolution can then be divided into two distinct phases: an initial Hawking-like phase where standard evaporation proceeds, and a subsequent “burdened” phase where quantum back-reaction effects slow down the mass loss. The transition between these phases occurs after a time \cite{Haque:2024eyh}
\begin{equation}
t_q = \tau_{\rm PBH}(1-q^3),
\end{equation}
here $q$ is the remaining fraction of the initial mass at the beginning of the burdened phase,
\begin{equation}
\label{eq:partmass}
M_{\rm PBH} = qM_{\rm PBH}^{i}.
\end{equation}
The choice of  $q$  only affects the way our results are interpreted in terms of the initial PBH mass $M_{\rm PBH}^{i}$. Throughout this Letter we assume  $q = 1/2$ , since memory burden effect is expected to become relevant after half the mass has evaporated. 

For  $t \geq t_q$ , quantum effects play a dominant role. The information accumulated on the PBH event horizon significantly damps the evaporation rate, introducing a dependence on the PBH entropy,  $S(M_{\rm PBH}) = 4 \pi G M_{\rm PBH}^2 $, which modifies the mass-loss equation as 
\begin{equation}
\label{eq:mbdot}
\frac{{\rm d}M_{\rm PBH}}{{\rm d}t} = \frac{1}{S(M_{\rm PBH})^{k}}\frac{{\rm d}M_{\rm PBH}^i}{{\rm d}t}, \quad k>0.
\end{equation}
Here $k$ parameterize the efficiency of the backreaction effect and is only constrained to be a positive number.
 Solving Eq. \ref{eq:mbdot}, we obtain the evolution of the PBH mass in the burdened phase
\begin{equation}
M_{\rm PBH}(t) = M_{\rm PBH} \left[ 1- \Gamma_{\rm PBH}^{(k)}(t-t_q)\right]^{1/(3+2k)} ,
\label{eq:Mtime}
\end{equation}
with
\begin{equation}
\Gamma_{\rm PBH}^{(k)} = \frac{\mathcal{G}\,g_{\rm H}}{7680 \pi}2^k(3+2k)M_P\left(\frac{M_{\rm P}}{M_{\rm PBH}}\right)^{3+2k} ,
\end{equation}
and $M_P=(8 \pi G)^{-1/2}$ being the reduced Planck mass. The total evaporation time is then
\begin{equation}
\tau_{\rm PBH}^{(k)} = t_q + (\Gamma_{\rm PBH}^{(k)})^{-1} \simeq (\Gamma_{\rm PBH}^{(k)})^{-1} .
\end{equation}
For  $k>0$ , this extended lifetime can be orders of magnitude longer than in the classical case.

\par This may have remarkable implications on the class of black holes believed to have been created in the very early Universe via the gravitational collapse of very large density inhomogeneities. This process occurs when an overdense region exceeding a critical density threshold, re-enters the horizon after inflation has ended~\cite{Carr:1975qj, Shibata:1999zs, Niemeyer:1999ak, Musco:2004ak}.

Due to the mechanism described, their mass function is almost unconstrained ``a priori'', and over the last decade this has caused consistent activity about the possibility that they could constitute all or a fraction of the Dark Matter, \cite{Cole:2023wyx} and references therein.
Constraints on this occurrence arise somewhat naturally for those PBHs which would have had to evaporate entirely by the Hubble time \cite{Boccia:2024nly,Luo:2020dlg,Kawasaki:1999na,Kawasaki:2000en,Kawasaki:2004qu,Kawasaki:2017bqm,Kawasaki:2020qxm,Hasegawa:2019jsa,deSalas:2015glj,Wu:2024uxa}, namely those whose masses don't exceed $M_{\rm PBH} \gtrsim 10^{15} \rm g$, whereas other types of physical mechanisms strongly constrain those masses that should survive the Universe's lifetime \cite{Carr:2020gox}.

The potential ``life-extending'' effect of memory burden with respect to a standard scenario has prompted a reassessment of the potential observations and of the constraints derived. It has also opened up to the enticing possibility that extremely high energy particles may be injected in the local Universe by the evaporation of PBHs of low masses, suppressed in the standard case \cite{Thoss:2024hsr,Chianese:2024rsn,Zantedeschi:2024ram,Loc:2024qbz}.

\par We study here the possibility that the neutrino claimed by the KM3NeT collaboration with the event KM3--230213A may be one of those particles.

\par{\bf Neutrino flux at Earth} We adopt a methodology similar to that used for our recent paper \cite{Chianese:2024rsn}, which addresses the flux of neutrinos emitted by a population of memory burdened PBHs, and its detectability within the IceCube detector.

We compute the neutrino flux expected by a population of PBHs constituting a fraction $f_{\rm PBH} = \rho_{\rm PBH}/ \rho_{\rm DM}$ of the total Dark Matter energy density. 

The three-flavour neutrino emission rate for a burdened PBH can be obtained by combining Eqs. \ref{eq:hawkspec} and \ref{eq:mbdot} with $g_i = g_{\nu} = 6$
\begin{equation}
    \frac{{\rm d}^2N_{\nu}}{{\rm d}E{\rm d}t} = S(M_{\rm PBH})^{-k}\frac{g_{\nu}}{2 \pi} \frac{\mathcal{F}(E,M_{\rm PBH})}{e^{E/T_{\rm H}} + 1} \,.
\end{equation}
In this study, we perform a numerical evaluation of the semi-classical neutrino emission rate using the \texttt{BlackHawk} code~\cite{Arbey:2019mbc,Arbey:2021mbl}. This framework also incorporates secondary neutrino production, modeled through the \texttt{HDMSpectra} hadronization process~\cite{Bauer:2020jay}. We assume a monochromatic mass spectrum for the PBHs in the considered mass range, and the emission spectrum is computed for the $M_{\rm PBH}$ mass today, as the emitting object is the one that has survived evaporation and not the initial one, as from Equation \ref{eq:partmass}.\\
If PBHs make up all or part of DM energy density, their spatial distribution should mirror that of DM. 
The galactic component of the flux is computed assuming a NFW density profile for the DM halo and performing the integral over the line of sight
\begin{equation}
    \frac{{\rm d}^2\phi^{\rm gal}}{{\rm d}E{\rm d}\Omega} =\sum_{\alpha} \frac{f_{\rm PBH}\,\mathcal{J}}{4 \pi M_{\rm PBH}} \frac{{\rm d}^2 N_{\nu_\alpha}}{{\rm d}E{\rm d}t}\,
    \label{eq:gal}
\end{equation}
here $\mathcal{J} = 2.22 \times 10^{22}~{\rm GeV/cm^2/sr}$ is the averaged J-factor and the sum is performed over the three neutrino families. We consider a generalized NFW density profile with scale radius of 25 kpc, a local DM density of $0.4~{\rm GeV/cm^3}$, and an index $\gamma=1$, values compatible with recent analysis~\cite{Iocco:2015xga,Benito:2019ngh,Benito:2020lgu}.
The extragalactic component is neglected, as we have verified that its contribution is subleading with respect to the contribution of our own Galaxy, \cite{Chianese:2024rsn}

The flux at Earth is then defined by:
\begin{equation}
    \phi_\nu(E) = \frac{{\rm d}^2\phi^{\rm gal}_{\nu}}{{\rm d}E{\rm d}\Omega}\,.
    \label{eq:flux_tot}
\end{equation}


\par{\bf Results} With the setup described so far, we compute the expected number of events in the whole detector energy range, for a given observation time $T$ as
\begin{equation}
n^{exp}(T) = 4 \pi \, T\int_{E_{\rm min}}^{E_{\rm max}} dE \, A_{\rm eff}(E)\, \phi_\nu(E) \, \, ,
\end{equation}
here $A_{\rm eff}(E)$ is the detector all-flavour sky-averaged effective area as reported in Ref.\cite{KM3NeT:2025npi}.

\begin{figure}[t]
    \centering
    \includegraphics[width=1.0\columnwidth]{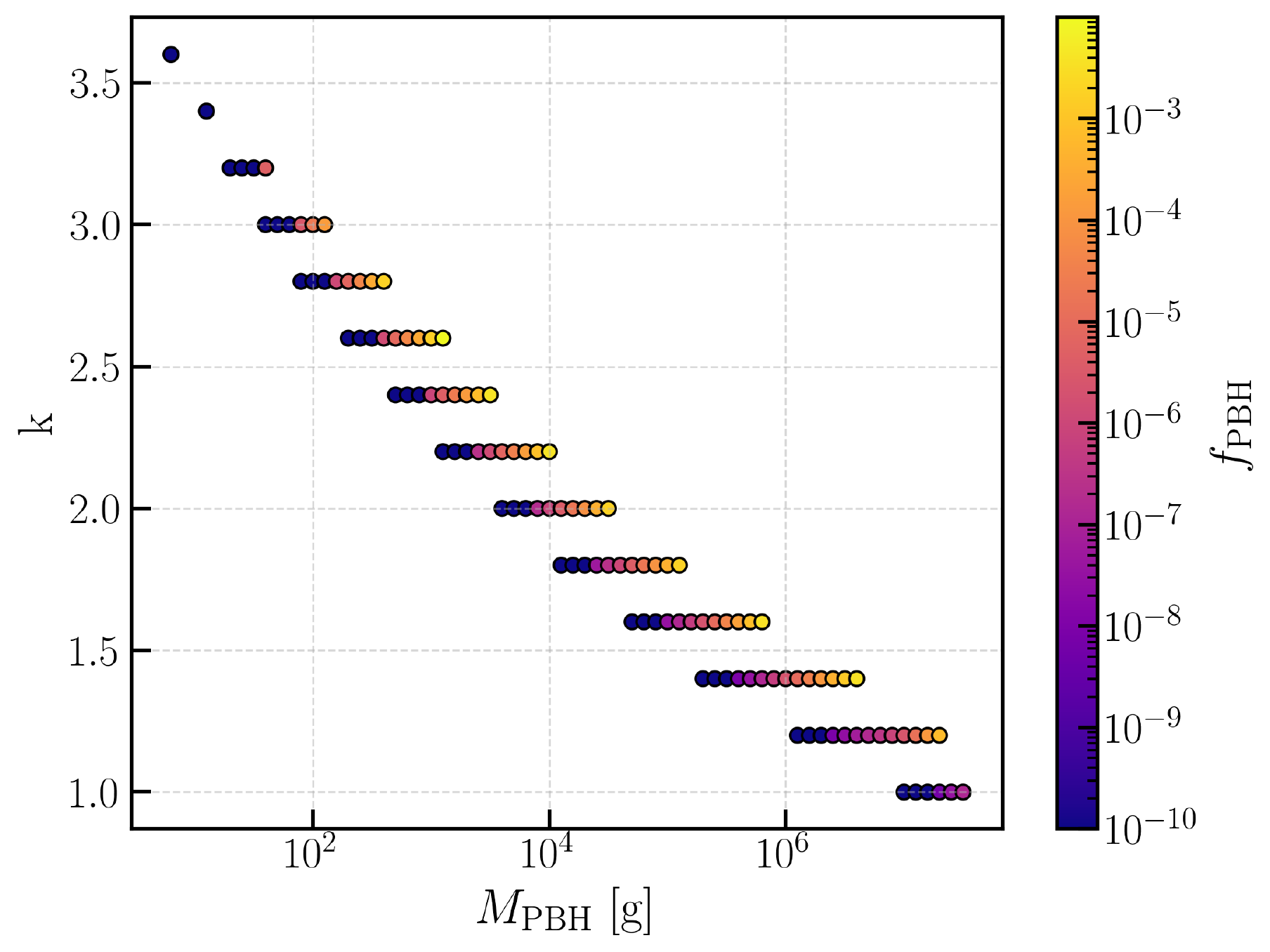}
    \caption{Candidates producing more than one event in the KM3NeT detector over a 6-month period with $f_{\rm PBH} = 1$. The color scale represents the maximum allowed $f_{\rm PBH}$ for these objects based on current constraints. }
    \label{fig:paramspacePBH}
\end{figure}

\begin{figure}[t]
    \centering
    \includegraphics[width=1.0\columnwidth]{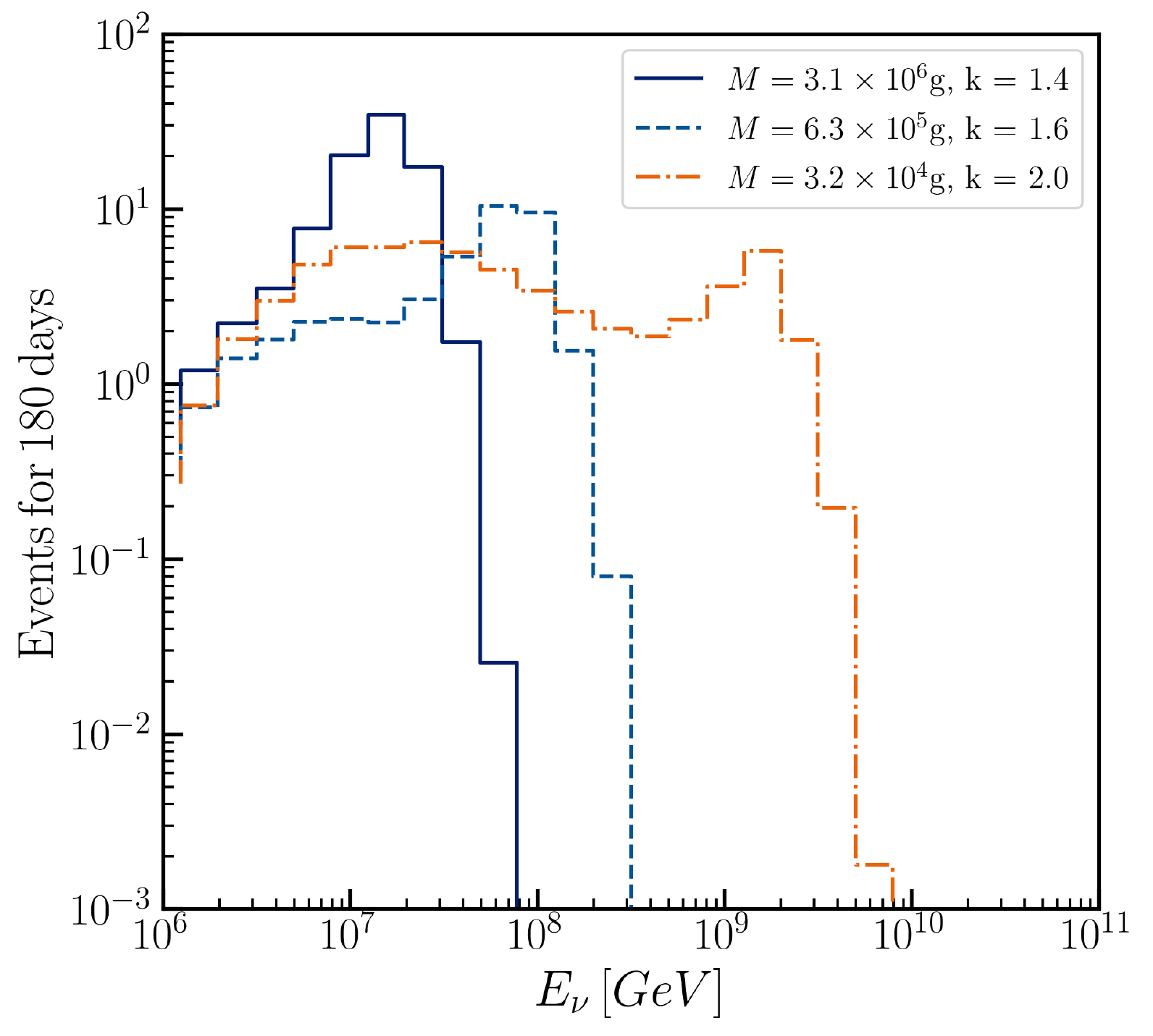}
    \caption{Expected number of neutrino events within KM3NeT per bin for three different ($M_{\rm PBH}, k$) combinations over a 180-day exposure. These candidates were selected by requiring that they produce at least one event, within the relevant energy range during this period, in the entire detector, assuming $f_{\rm PBH} = 1$.}
    \label{fig:candidates}
\end{figure}

We scan the ($M_{\rm PBH}- k$) parameter space of memory-burdened PBHs, computing the neutrino flux expected at Earth, and identifying those configurations (PBHs with mass and k such that) are able to produce at least one neutrino event in a six month time (180 days) within the entire KM3NeT detector, under the assumption that they constitute the entirety of the Dark Matter.\\

Before introducing our results, it is important to stress that the setup described until now served only the purpose to identify the ballpark parameter space, and that several caveats are in order: {\it a)} for some of the $k$ identified, PBHs lifetimes are shorter than the current age of the Universe. We discard those configurations; {\it b)} we are not using the entire exposure time for the event observed, we discuss this later; {\it c)} constraints of different nature on $f_{\rm PBH}$ exist, we address this in the following. The ($M_{\rm PBH}, k$) combinations thus identified are reported in the Figure \ref{fig:paramspacePBH}, with the distribution of expected events shown in Figure \ref{fig:candidates} for those which produce the highest flux at Earth.

\begin{table}[h]
    \centering
    \renewcommand{\arraystretch}{1.3} 
    \setlength{\tabcolsep}{8pt}      
    \begin{tabular}{c c c c c}
        \hline
        $M_{\mathrm{PBH}}\,[\mathrm{g}]$ & $k$ & $f_{\mathrm{PBH}}$ & $t_{\mathrm{KM}}\,[\mathrm{yr}]$ & $t_{\mathrm{IC2}}\,[\mathrm{yr}]$ \\
        \hline
        $2.0 \times 10^6$  & 1.4 & $2.1 \times 10^{-5}$      & $21$  & $5.2 \times 10^{-1}$ \\
        $2.0 \times 10^6$  & 1.6 & $8.6 \times 10^{-1}$      & $21$  & $5.2 \times 10^{-1}$ \\
        $4.0 \times 10^6$  & 1.2 & $2.8 \times 10^{-8}$      & $19$  & $3.7 \times 10^{-1}$ \\
        $4.0 \times 10^6$  & 1.4 & $1.5 \times 10^{-3}$      & $19$  & $3.7 \times 10^{-1}$ \\
        $5.0 \times 10^6$  & 1.2 & $1.6 \times 10^{-7}$  & $12$  & $2.1 \times 10^{-1}$ \\
        $5.0 \times 10^6$  & 1.4 & $8.7 \times 10^{-3}$      & $13$  & $2.2 \times 10^{-1}$ \\
        $6.3 \times 10^6$  & 1.2 & $3.6 \times 10^{-7}$  & $20 $  & $2.9 \times 10^{-1}$ \\
        \hline
    \end{tabular}
    \caption{Exposure time for one neutrino event $t_{\mathrm{KM}}$ to be expected within the current KM3NeT configuration for the most highest-yielding pairs of $(M_{\mathrm{PBH}}, k)$, assuming they constitute the highest dark matter fraction $f_{\mathrm{PBH}}$ allowed by current constraints. We also report forecast times for one neutrino event to be expected within the forthcoming IceCube Gen-2 detector, $t_{\mathrm{IC2}}$.}
    \label{tab:top10_events}
\end{table}

\par {\bf Discussion and forecasts} In light of the caveats introduced, we note at this point that existing constraints limit $f_{\rm PBH} \ll 1$ for all the parameter combinations found. Namely: none of the pairs of $M_{\mathrm{PBH}}$ and $k$ found to produce a sizable flux can form the entirety of the dark matter, and the number counts shown in the plot must be appropriately rescaled for each point in Figure \ref{fig:paramspacePBH}. This leads --for all combinations identified, for the actual exposure time of the event observed (335 days), and for the directional characteristics of the event-- to an expected event number smaller than one.

Fully aware of the above --yet mindful that the year 2025 has already delivered its fair share of discouraging news-- we choose to postpone any further negative statements, adopting what some of our colleagues would call a ``positivist'' perspective \cite{Carr:2023tpt} on what others have defined as a ``strike of luck'' \cite{Scotti25}.

We assume that the KM3--230213A event has indeed been produced by a neutrino emitted by an evaporating black hole of low mass, survived until today owing to the memory-burden effect. Within this scenario, it is now possible to make predictions of the number of future events expected within the KM3NeT detector in the future, for each mass $M_{\mathrm{PBH}}$ and $k$ identified, and assuming that their contribution to the dark matter fraction is maximal, i.~e.~ the highest $f_{\rm PBH}$ allowed by the current constraints. 
The latter have been obtained by the absence of neutrino events with the correct characteristics within the IceCube detector in a seven year dataset \cite{Chianese:2024rsn}, and that of $\gamma$-rays within several $\gamma$ observatories \cite{Thoss:2024hsr,Chianese:2025wrk}. By adopting the values of $f_{\rm PBH}$ shown in Table \ref{tab:top10_events} into Eq.\ref{eq:gal}, we are ensuring that our working hypothesis does not violate the lack of previous observations.
It is worth noting here that the tension between the KM3NeT and the IceCube data found by \cite{Li:2025tqf} does not apply at face value to our scenario, as it has been derived for cosmogenic neutrinos and an isotropic signal. This is not the case for the PBH--induced neutrino flux, which follows a Galactic DM distribution.

We apply the methodology described above to perform this analysis, presenting the results in Table \ref{tab:top10_events}. We show there the combinations ($M_{\mathrm{PBH}}$, $k$) with the highest expected event counts, assuming that each specific pair of black hole mass $M_{\mathrm{PBH}}$ and memory burden strength $k$ contributes the maximum permissible fraction of Dark Matter $f_{\mathrm{PBH}}$ within existing constraints, and report the corresponding exposure time required to expect one neutrino event within the KM3NeT. 
It is to be noted here that the configuration of the experiment is constantly changing as new detectors are added, with the effect of enhancing the effective area with a rate that is not possible to predict reliably. Our estimates are therefore to be regarded as conservative, and to be updated as the experiment keeps being upgraded. Given these uncertainties, we refrain from presenting more detailed detection forecasts --such as the angular dependence of the signal due to the orientation of the densest dark matter regions in the Galaxy, or exclusion plots based on predicted sensitivities-- as they fall beyond the scope of this study. A more comprehensive study should be undertaken once the updated instrument configurations become available.
We also report forecasts for the exposure time in the IceCube Gen-2 detector  --currently under construction, and with effective area estimated as from \cite{IceCube-Gen2:2020qha}-- to show the potential of future generation detectors. Also, gamma-ray observations should accompany future neutrino ones, so specific forecasts should be performed in order to test, or constrain, the hypothesis put forward in this paper via an appropriate multi-messenger strategy for the entire memory burdened PBH parameter space.

\par{\bf Conclusions} Within a cosmological framework where primordial black holes (PBHs) constitute a fraction of dark matter ($f_{\rm PBH}$), we have investigated whether PBHs evaporating today could have produced the KM3–230213A event by emitting a spectrum of particles, including neutrinos. For this scenario to be viable, their lifetime must be extended by the recently proposed memory-burden effect.  We have identified the PBH masses $M_{\mathrm{PBH}}$ and the memory burden effect strength parameter $k$ that maximize the number of expected events in the KM3NeT detector while remaining consistent with existing observational constraints. Although the probability that KM3–230213A was produced by an evaporating memory-burdened PBH is very low ($\ll 1$) for all identified parameter sets ($M_{\mathrm{PBH}}$, $k$) identified, one can nevertheless take an optimistic stance and put this hypothesis to the test.

Following this approach, we predict the number of neutrino events expected in KM3NeT and Icecube Gen-2 under the assumption that PBHs contribute the maximum fraction of dark matter $f_{\rm PBH}$ allowed by current constraints. The predicted event rate is significant, reaching the order of unity within a few years. This would make it possible to either confirm or rule out this scenario, subjecting the memory-burdened PBH paradigm to imminent scrutiny with neutrino observations. \\

{\bf Acknowledgements} We acknowledge support by the research project TAsP (Theoretical Astroparticle Physics) funded by the Istituto Nazionale di Fisica Nucleare (INFN). The work of FI is further supported by the research grant number 2022E2J4RK ``PANTHEON: Perspectives in Astroparticle and Neutrino THEory with Old and New messengers'' under the program PRIN 2022 funded by the Italian Ministero dell’Università e della Ricerca (MUR).

\bibliography{bibliography}
\end{document}